\documentclass[prl,twocolumn,floatfix]{revtex4}

\usepackage[dvips]{graphicx}
\usepackage{amssymb}
\usepackage{amsmath}

\newcommand{\be}{\begin{equation}}
\newcommand{\ee}{\end{equation}}

\begin{document}

\title{
\vskip -80pt
{\begin{normalsize}
\mbox{} \hfill DAMTP-2006-104 \\
\vskip 10pt
\end{normalsize}}
{\bf\Large
Kac--Moody Algebras and Controlled Chaos
} }


\author{Daniel H. Wesley}\email{D.H.Wesley@damtp.cam.ac.uk}

\affiliation{Department of Applied Mathematics and Theoretical Physics,
  Centre for Mathematical Sciences, University of Cambridge,
  Wilberforce Road, Cambridge CB3 OWA, United Kingdom}

\date{4 November 2006}

\begin{abstract}
\noindent Compactification can control chaotic Mixmaster behavior in
gravitational systems with $p$--form matter: we consider this in light
of the connection between supergravity models and Kac--Moody algebras.
We show that  different
compactifications define ``mutations'' of the algebras associated with
the noncompact theories.  We list the algebras obtained in this way,
and find novel examples of wall systems determined by Lorentzian (but
not hyperbolic) algebras.  
Cosmological models with a smooth pre--big bang
phase require that chaos is absent: we
show that compactification alone cannot
eliminate chaos in the simplest compactifications of the heterotic
string on a Calabi--Yau, or M theory on a manifold of $G_2$ holonomy. 
\end{abstract}

\maketitle 


  \section{Introduction}


It has long been known that gravitational systems typically approach a big
crunch chaotically -- a phenomenon sometimes called
``Belinskii--Khalatnikov--Lifshitz (BKL) oscillations" or 
``Mixmaster behavior" \cite{mixmaster}.  
Spacetime becomes highly
anisotropic,  and the directions and rates of contraction (and
expansion) oscillate chaotically until the big crunch is reached.
This phenomenon deserves study for several reasons.
For one, it is quite common 
among suitably general solutions of the Einstein equations near
spacelike singularities.  Also, the study of supergravity models in this
regime has revealed algebraic structures that are related to a
conjectured underlying U--duality symmetry group of M theory \cite{e10}.
Lastly, avoiding chaotic behavior is essential for any 
cosmological model with a pre--big bang
phase; the universe must enter the 
expanding era in a nearly
isotropic and homogeneous state, which seems unlikely after
a chaotic epoch.

Here we characterize the chaotic properties
of the low--energy supergravities obtained from simple compactifications
of the heterotic string and M theory.
We begin with the  remarkable fact that the dynamics of these theories in
the BKL limit  is controlled by the 
$BE_{10}$ and $E_{10}$ Kac--Moody algebras \cite{billiards}.  This
structure is invariant under toroidial compactification
\cite{ox-red}, but 
non--toroidial compactification
changes the standard picture \cite{Wesley:2005bd}
by deleting and adjoining roots of these
algebras according to
simple rules.  Each compactification thereby produces a ``mutation'' of the
original algebra.  
We apply this fact and obtain novel examples where the dynamics is
controlled by Lorentzian algebras that are not hyperbolic, 
previously seen only in Einstein gravity in
spacetime dimension $\ge 11$ or when the  noncompact spectrum is
constrained as in \cite{geometric}.
We also list the compactifications for which chaos is suppressed: 
these provide a set of potential cosmological models 
whose pre--big bang phases are free of chaos.

In the present work, we consider only ``simple''
Kaluza--Klein compactifications: quantum effects,
fluxes, D-- or M--branes, orbifolds, conifolds, etc., are not 
included, though it
would be interesting to discover how these affect our conclusions.
We also work in a
free field approximation, but even if potentials are included, 
the only possibility relevant for chaos appears to be
a matter component with $P > \rho$ \cite{Erickson:2003zm},
as required by the cyclic universe
\cite{cyclic}.  Most other interactions
\cite{Damour:2002tc}, as well as matter with $P < \rho$, are 
irrelevant near a big crunch.


  \section{The Wall System}\label{s:notation}


We first review some essential facts regarding the wall systems
corresponding to gravitational theories,
following \cite{billiards}.  We study 
 cosmological spacetimes with string frame metric
\be
\mbox{d}s^2 = 
-N^2 e^{-2\beta^0(t)} \mbox{d}t^2 + 
\sum_{i=1}^d e^{-2\beta^i(t)} [\omega^i]^2,
\ee
in which spatial curvature is included through the choice of 
$\omega^i = {\omega^i}_j(x) \,\mbox{d}x^j$, and $d=9$ or $10$, 
depending on whether we are studying a
ten dimensional superstring theory or eleven dimensional
supergravity.  In the superstring case, we define
$\beta^0(t) = \sum_{i=1}^9 \beta^i(t) +  2\Phi(t) $,
and fix $N=n \exp{\left(-\beta^0\right)}$, while for the M theory case 
$N=n \exp{\left(-\sum_{j=1}^{10}\beta^j\right)}$.  
In either case, we will refer to
the variables $\beta^\mu$ with  $\mu=0,\dots,9$ or
$1,\dots,10$ as required.  The spacetime action becomes 
\be\label{eq:beta_action}
S = \int \left( \frac{\eta_{\mu\nu}}{2n} 
\frac{d\beta^\mu}{dt} \frac{d\beta^\nu}{dt}
- nV(\beta) \right) \; \mbox{d}t ,
\ee
where $\eta_{\mu\nu}$ is a flat metric of signature
$(-+\cdots+)$.
 The effects of $p$--form energy densities, spatial gradients, and
curvature in the physical spacetime are thereby encoded in the 
Toda--type potential 
\be\label{eq:V_beta}
V(\beta) = \sum_A |c_A| \exp{\left( -2w_{A\mu} \beta^\mu \right)} 
\ee
for the motion of the point $\beta^\mu(t)$ in an auxiliary spacetime.
The $\{w_{A}\}$ are ``wall forms,'' indexed by $A$ and with components
$w_{A\mu}$. They are determined by the $p$--form menu,
 and each corresponds to a 
Kasner stability condition
 \cite{mixmaster,Wesley:2005bd,billiards}.
The coefficients $c_A$ depend on the initial energy densities
in $p$--form fields, curvature perturbations, spatial gradients, and
other contributions.  

It is conventional to simplify matters by restricting our
attention to a smaller set $\{r_A\} \subset \{w_A\}$ of 
``dominant walls,'' which are not hidden behind other walls.
In the case of superstrings and M theory, the Cartan matrix
$A_{AB} = 2 (r_A \cdot r_B)/(r_A \cdot r_A)$,
computed using the natural metric $\eta_{\mu\nu}$ on the $\beta$--space,  
is precisely that of the $E_{10}$ and $BE_{10}$ Kac--Moody algebras 
\cite{billiards}.  
The dominant walls play the role of 
simple roots of the algebra.  For the $E_{10}$ and $BE_{10}$ cases,
the point $\beta^\mu(t)$ is trapped by the dominant
walls, and so the
corresponding string and M theory models have only chaotic solutions,
undergoing an infinite number of BKL oscillations as they approach the
big crunch.


  \section{Compactification and Mutation}\label{s:meat}


Compactification changes the wall system associated with a given
theory.  To see why, consider that when spacetime is non--compact,
each $p$--form posseses a spatially homogeneous mode
that can grow rapidly near the big crunch.  These modes correspond to
dominant walls and are responsible for chaos.   The wall system is 
unchanged by toroidial compactification \cite{billiards,ox-red}, but 
it was shown in \cite{Wesley:2005bd} that compactification on more 
general manifolds can avoid chaos by forbidding the spatially
homogeneous modes of some $p$--form fields.  The energy density
in the remaining modes scales like dust or radiation, which are
irrelevant during a contracting phase.  

The influence of compactification on the wall system is expressed by a
``selection rule'' \cite{Wesley:2005bd}:
If a Betti number $b_j$ of the compact manifold 
vanishes, then remove from $\{w_A\}$ all $p$--form walls arising from
the electric modes of a $j$--form or the magnetic modes of a
$(j-1)$--form.
(We take the ``$p$'' in ``$p$--form'' to be the number of indices on
its gauge potential.)
There is a corresponding rule for gravitational walls, which we will
not 
 employ here.
The selection rules are subject to a genericity assumption  that the compact 
manifold $M$ does not factor, both topologically and metrically,
as $M=M_1 \times M_2$.

Since compactification deletes walls, it modifies the dominant wall set.
If the Cartan matrix for the new dominant wall set
obeys the generalized Cartan conditions \cite{Kac}, it 
defines a new algebra, the ``mutation'' of the original algebra.
There is no {\em a priori} reason that special properties of the
dominant wall set should survive compactification, and in the next
section we will give examples where the new wall systems do not define
an algebra.

A natural question arises: can compactification remove enough walls to
allow non--chaotic solutions to a previously chaotic theory, such as
string or M theory?  This was addressed in 
\cite{Wesley:2005bd}, where some non--chaotic solutions were found.
Here we will give a more complete answer, using the ``coweights" 
$\Lambda^{\vee A}$  introduced in \cite{billiards}.  If the $\{r_A\}$ are
linearly independent and complete, the coweights are defined by the 
condition
$r_{A\mu} \Lambda^{\vee B\mu} = {\delta_A}^B$.
The region far from the walls is then given by the cone $W^+$ of
linear combinations of coweights with nonnegative coefficients.
Non--chaotic solutions of (\ref{eq:beta_action}) are null rays whose
velocities lie in $W^+$.  These solutions exist only when there are
both spacelike and timelike coweights; 
thus proving the existence of non--chaotic solutions reduces to
computing the norms of the coweights.


  \section{Heterotic String and M Theory}\label{s:het}


\begin{figure}
  \begin{center}
    \includegraphics[width=3.4in]{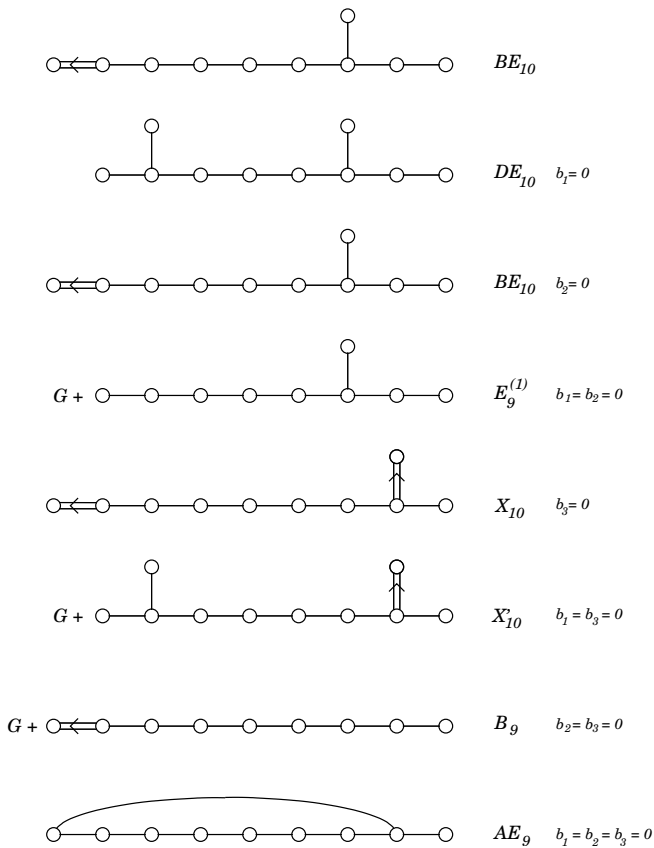}
  \end{center}
  \caption{The root systems arising from the compactification of the 
           heterotic string on a six--manifold.  There are two unnamed rank ten
           diagrams, denoted $X_{10}$ and $X'_{10}$.
           ``$G$'' denotes a gravitational wall. }
  \label{f:het_dynkins}
\end{figure}

\begin{table}
  \begin{tabular}{|c|c|c|c|c|c|c|c|}
    \hline      
     \multicolumn{3}{|c|}{Betti numbers} 
      & chaotic? &
     \multicolumn{3}{|c|}{num. of $\Lambda^\vee$ that are:} & ``root \\
    $\enspace b_1\enspace$  & $\enspace b_2\enspace$ & 
    $\enspace b_3\enspace$ &  & $\enspace$s--like$\enspace$ & null 
    & $\enspace$t--like $\enspace$
    & $\enspace$ system" $\enspace$\\
     \hline
      + & + & + & yes & 0 & 2 & 8 & $BE_{10}$ \\
      0 & + & + & yes & 0 & 3 & 7 & $DE_{10}$ \\
      + & 0 & + & yes & 0 & 2 & 8 & $BE_{10}$ \\
     \hline
      0 & 0 & + & no  & 1 & 8 & 1 & $G+E_9^{(1)}$ \\
      + & + & 0 & no  & 5 & 2 & 3 & $X_{10}$ \\
      0 & + & 0 & no  & - & - & - & $G + X'_{10}$\\
      + & 0 & 0 & no  & 5 & 2 & 3 & $G+B_9$ \\
      0 & 0 & 0 & no  & 1 & 1 & 8 & $AE_9$ \\
      \hline
    \end{tabular}
    \caption{Summary of the compactifications of the heterotic theory
             on $M_6$.  Vanishing Betti numbers are denoted by a
             ``0,'' non--vanishing ones with a ``+.''
             ``$G$" denotes a gravitational wall form that, when
             deleted, leaves a remaining root system of Lie type. Two
             previously unnamed root systems appear, denoted $X_{10}$
             and $X'_{10}$.  The $b_1=b_3=0$ examples possess special
             properties as described in the text.}
      \label{t:het_sum}
  \end{table}

We now possess the tools required to describe
the chaotic properties of a theory, given its $p$--form menu 
 and the Betti numbers of its compactification
manifold, which we now apply to the supergravities obtained
from the heterotic string and M theory.
For each theory,
we begin with the full set of
billiard walls as given in \cite{billiards,ox-red}.  We then remove walls in
accordance with the selection rules described in the previous section
and in \cite{Wesley:2005bd}, and find the new relevant walls.  From
these we calculate the new coweights and new Cartan matrices, which
determine the chaotic properties of the compactified theories as per
our discussion above.

The results for the heterotic string 
are summarized in Figure \ref{f:het_dynkins} and
Table \ref{t:het_sum}. We assume compactification on a six-manifold,
which has  three relevant Betti numbers
$b_1$, $b_2$ and $b_3$, and therefore eight possible 
vanishing Betti number combinations.
Our results indicate that controlled chaos requires
either $b_3$ or both $b_1$ and $b_2$ to vanish.  This agrees with
numerical searches for Kasner solutions with
controlled chaos carried out by the author.  
The formulation reported herein is superior to
that of \cite{Wesley:2005bd} and the numerical search since we can say 
definitively that these are the {\em only} solutions with controlled
chaos.  Significantly, Calabi--Yau compactifications, which have 
$b_1 = 0$ but $b_2,b_3 >0$, appear incompatible with controlled chaos.

A wide variety of root systems arise from compactifications of the heterotic
string. While the original
wall system is described by a Dynkin diagram, 
one sees from Figure \ref{f:het_dynkins} that this property is not shared
by all of the wall systems after compactification.  However, a valid
Dynkin diagram is always obtained by omitting a single gravitational wall.
The diagrams so obtained include algebras of 
           Lorentzian ($X_{10}$, $X'_{10}$), 
           hyperbolic ($BE_{10}$, $DE_{10}$,
           $AE_9$), affine ($E^{(1)}_9$), and finite ($B_9$) type.

Compactification provides novel examples of
systems where the dynamics is controlled by algebras
that are Lorentzian but not hyperbolic -- {\em i.e.}
the simple root Gram matrix has signature
$(-+\cdots +)$, but deleting any single node does {\em not} yield a
finite or affine algebra.
 The relevant new algebras are 
$X_{10}$ and $X'_{10}$ in Figure \ref{f:het_dynkins}.
They are examples of the
 ``very extended'' Lie algebras of the type studied
in \cite{very-over-x}, with various possible central node assignments.
The appearance of these algebras is notable, since 
previous examples of algebras arising in the BKL limit have all been
hyperbolic,
except for the $AE_d$ series with $d \ge 10$ for pure gravity in
$(d+1)$ dimensions, and some Lorentzian algebras obtained by
geometric constraints on M theory fields
\cite{geometric}.

Special care is required when all three Betti numbers vanish, for 
the relevant wall set is linearly 
independent but incomplete.  We can recover the coweight description
by introducing a (spacelike) covector
${\widehat \Lambda}^{\vee}$ such that
$r_{A\mu} {\widehat \Lambda}^{\vee\mu} = 0$ for all $A$.
The $\Lambda^{\vee A}$ are then determined up to 
$\Lambda^{\vee A} \sim \Lambda^{\vee A} + k^A {\widehat
\Lambda}^{\vee}$ for constants $k^A$, which we fix  by 
requiring
$\eta_{\mu\nu} \Lambda^{\vee A \mu} {\widehat \Lambda}^{\vee\nu} = 0$,
which amounts to minimizing the norm of the coweights. 
Now by analogy to the usual case we can
treat ${\widehat \Lambda}^{\vee}$ as a coweight and
 write points in $W^+$ as
$\widehat \lambda  {\widehat \Lambda}^{\vee} +
 \sum_A \lambda_A  \Lambda^{\vee A}$,
with $\lambda_A \ge 0$.  Since
$\widehat \lambda$ may take any value,
 $W^+$ is no longer a cone.  Nonetheless, ${\widehat \Lambda}^\vee$ is
spacelike and therefore non--chaotic solutions exist.
When only $b_2 >0$, the dominant walls are not linearly independent,
and so a coweight description is impossible: however it is clear that
this compactification has non--chaotic solutions since the $b_3=0$
compactification does.

\begin{figure}
  \begin{center}
    \includegraphics[width=3.4in]{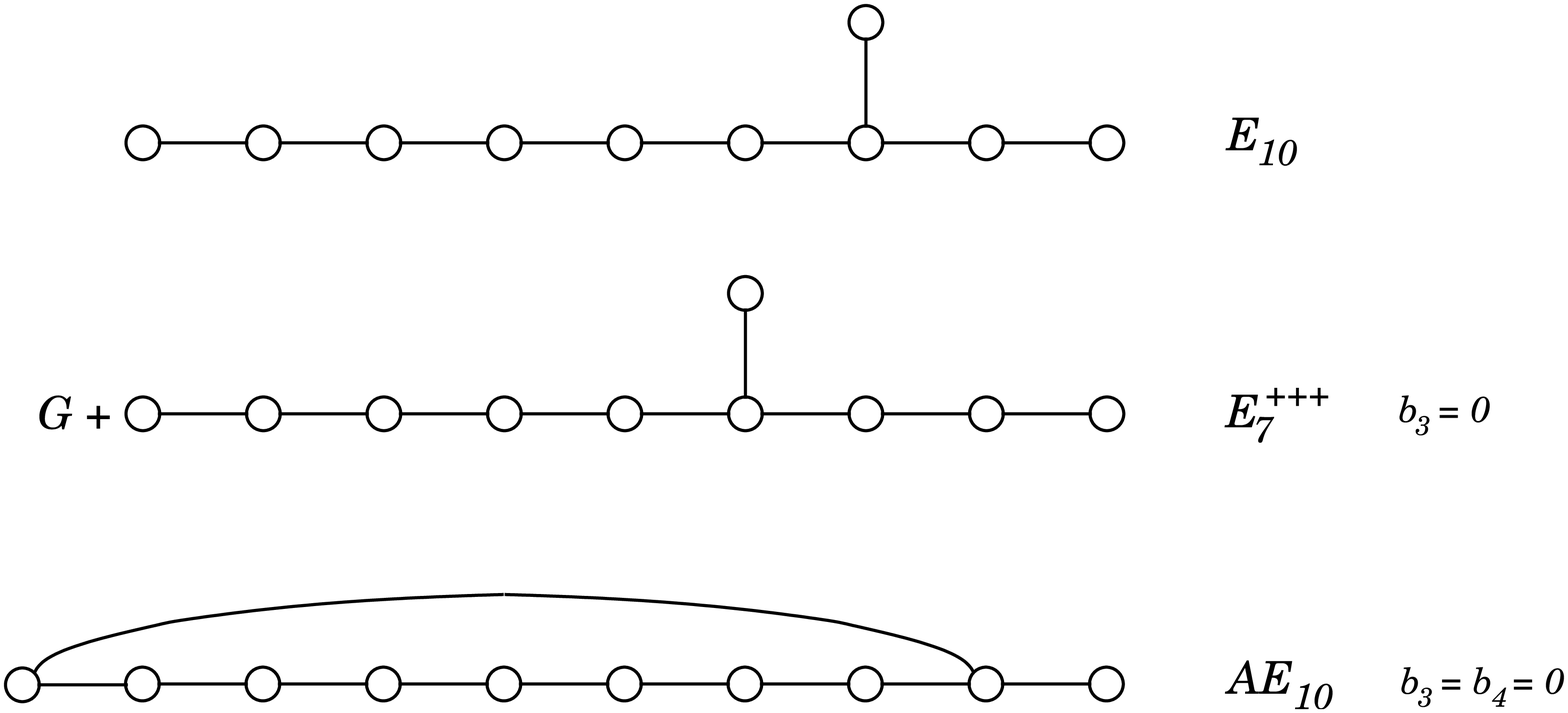}
  \end{center}
  \caption{The Dynkin diagrams appearing in the compactification of
           M theory.  }
  \label{f:M_dynkins}
\end{figure}

We next 
consider
eleven dimensional
supergravity.  The results are shown in Figure
\ref{f:M_dynkins}.  The chaotic properties are entirely controlled by
two Betti numbers, $b_3$ and $b_4$, and
in compactifications to four
dimensions we have $b_3 = b_4$ by Poincar\'{e} duality.  Therefore the
only relevant algebras for compactification to four dimensions are 
$E_{10}$ (when $b_4 > 0$) and 
$AE_{10}$ (when $b_3=b_4=0$).  $E_7^{+++}$ was found to control the
dynamics of an M theory truncation
 in \cite{geometric}, though via a different construction and
without the additional
gravitational wall present here.

Our results imply that chaos can be controlled if $b_3 = b_4 = 0$.
This yields $AE_{10}$, which is Lorentzian (but not
hyperbolic) and possesses spacelike coweights.
This is expected:
with this compactification the massless bosonic sector is identical to
vacuum Einstein gravity, whose billiard system is described by
$AE_{10}$ and is not chaotic \cite{mixmaster}.


  \section{Conclusions}\label{s:conclusions}


We have studied how compactification influences the algebraic structures
controlling the dynamics of string and M theory models near a big crunch.
Different compactifications of the heterotic string and M theory,
defined by their vanishing Betti numbers, lead to
``mutations'' of the $BE_{10}$ and $E_{10}$ algebras that control the
BKL dynamics of the noncompact theories.
We have listed the set of algebras obtained
from simple compactifications, and have found new examples where the
BKL dynamics is controlled by 
Lorentzian (but not hyperbolic)
algebras, denoted here by $X_{10},X'_{10}$, and $E_7^{+++}$.
Previously, algebras in this class have only been observed controlling the
BKL dynamics of pure Einstein gravity in
$(d+1) \ge 11$ dimensions, where the $AE_d$ series appears, or in
\cite{geometric}.

Our results rule out
controlling chaos in the sense of \cite{Wesley:2005bd} within
the simplest string models of four--dimensional physics.
For the heterotic string compactified on a 
Calabi--Yau manifold \cite{Candelas:1985en},
 the first Betti number
$b_1 = 0$, but both $b_2$ and $b_3$ are nonzero. As
Table \ref{t:het_sum} indicates, compactification with only $b_1$ vanishing is
insufficient to control chaos.  A similar problem arises with M
theory, compactified on a seven--manifold of $G_2$ holonomy
\cite{G2}.  In this
case $b_3$ and $b_4$ are nonzero, and we have shown here that 
chaos is inevitable for a compactification with these Betti numbers.

Our work connects with other results regarding modifications
of the algebras that control BKL dynamics.  Lorentzian subalgebras 
of $E_{10}$
were
uncovered in \cite{geometric} through various truncations of the
spectrum of M theory.  In \cite{ox-red} 
 chains of theories controlled by the same (hyperbolic) algebra are 
obtained by 
by dimensional 
oxidation and reduction.  The results presented here indicate that it is 
possible to jump between chains through oxidation or reduction on manifolds
with vanishing Betti numbers.

It would be interesting to understand how the selection rules are
modified in more general 
string and M theory compactifications (including flux, 
D-- or 
M--branes, orbifolds, etc. -- see \cite{further}).
These wider classes of string models hold the
promise of providing a realistic low energy particle spectrum, and we
expect that the techniques employed here would be useful for determining
whether controlled chaos is possible.
It would be significant for cosmology
if a satisfactory compactification could be found,
as it might form the basis of a new
string cosmological model with a smooth pre--big bang phase.


  \section{Acknowledgements}


 We are grateful to Marc Henneaux and Daniel Persson for many useful comments,
 assistance with 
 terminology, identifying the $E_7^{+++}$ algebra, and for
 pointing out a number of interesting references.
 We enjoyed some informative conversations with
 Malcolm Perry and Neil Turok
 regarding Kac--Moody
 algebras and their significance in physics  
 during the later stages of this work. 
 We thank Katie Mack for a
 careful reading of this manuscript, and Latham Boyle and
 Andrew Tolley for their comments on an earlier version.


\end{document}